\documentclass[a4paper]{jpconf}
\usepackage{citesort}

\usepackage[pdftex]{graphicx}

\usepackage[tbtags]{amsmath} % assumes amsmath package installed
\usepackage[cmintegrals]{newtxmath}
\usepackage{array}
\usepackage{amssymb}
\usepackage{url}
\usepackage{siunitx}
\NewDocumentCommand\angRange{O{} m m}{\SIrange[parse-numbers=false,#1]{\ang[parse-numbers=true]{#2}}{\ang[parse-numbers=true]{#3}}{}}
\usepackage{tabulary}
\usepackage{tabularx}
\usepackage[font={small}]{caption}
\usepackage[font={small}]{subcaption}
\usepackage{balance}
\usepackage{nicefrac}
\usepackage{comment}

\usepackage{cuted}
\allowdisplaybreaks[1]

\usepackage[algo2e,vlined,ruled]{algorithm2e}

 \usepackage{tikz,pgfplots}
 \usetikzlibrary{spy}
 \pgfplotsset{compat=newest}
 \newlength\figureheight
 \newlength\figurewidth
 \pgfkeys{
         /pgfplots/scaled ticks = true,
         /pgf/number format/sci subscript
    }
 \pgfplotsset{plot coordinates/math parser=false}
 \pgfplotsset{every axis/.append style={scaled x ticks = false}}

\usetikzlibrary{external}
\begin{document}
\def\xku{\mathbf{\hat{x}}_k^u}
\def\xkuu{\mathbf{\hat{x}}_{k+1}^u}
\def\xkp{\mathbf{\tilde{x}}_k^p}
\def\xkpp{\mathbf{\tilde{x}}_{k+1}^p}
\def\xkc{\mathbf{\tilde{x}}_k^c}
\def\xkcc{\mathbf{\tilde{x}}_{k+1}^c}
\def\xke{\mathbf{\tilde{x}}_k^e}
\def\xkee{\mathbf{\tilde{x}}_{k+1}^e}
\def\bA{\mathbf{A}}
\def\bB{\mathbf{B}}
\def\bC{\mathbf{C}}
\def\bD{\mathbf{D}}
\def\bI{\mathbf{I}}
\def\bW{\mathbf{W}}
\def\bM{\mathbf{M}}
\def\bP{\mathbf{P}}
\def\bU{\mathbf{U}}
\def\bY{\mathbf{Y}}
\def\bL{\mathbf{L}}
\def\bV{\mathbf{V}}
\def\bK{\mathbf{K}}
\def\bX{\mathbf{X}}
\def\bR{\mathbf{R}}
\def\bQ{\mathbf{Q}}
\def\bSigma{\mathbf{\Sigma}}
\def\bLambda{\mathbf{\Lambda}}
\def\bu{\mathbf{u}}
\def\ba{\mathbf{a}}
\def\by{\mathbf{y}}
\def\bz{\mathbf{z}}
\def\bv{\mathbf{v}}
\def\br{\mathbf{r}}
\def\bs{\mathbf{s}}
\def\bl{\mathbf{l}}
\def\bw{\mathbf{w}}
\def\bx{\mathbf{x}}
\def\bm{\mathbf{m}}
\def\bk{\mathbf{k}}
\def\br{\mathbf{r}}
\def\bd{\mathbf{d}}
\def\bc{\mathbf{c}}
\def\bff{\mathbf{f}}
\def\calN{\mathcal{N}}
\def\bmu{\boldsymbol{\mu}}
\def\bnu{\boldsymbol{\nu}}
\def\bphi{\boldsymbol{\phi}}
\def\bpsi{\boldsymbol{\psi}}
\def\btau{\boldsymbol{\tau}}
\def\bPhi{\boldsymbol{\Phi}}
\def\bomega{\boldsymbol{\omega}}
\def\blambda{\boldsymbol{\lambda}}
\def\bLambda{\boldsymbol{\Lambda}}
\def\bSigma{\boldsymbol{\Sigma}}
\def\bGamma{\boldsymbol{\Gamma}}
\def\bPsi{\boldsymbol{\Psi}}

\title{Adaptation of Engineering Wake Models using Gaussian Process Regression and High-Fidelity Simulation Data}

\author{Leif Erik Andersson$^{\boldsymbol{\mathsf{1}}}$, Bart Doekemeijer$^{\boldsymbol{\mathsf{2}}}$, Daan van der Hoek$^{\boldsymbol{\mathsf{2}}}$, Jan-Willem van Wingerden $^{\boldsymbol{\mathsf{2}}}$, Lars Imsland$^{\boldsymbol{\mathsf{1}}}$}

\address{$^{\mathsf{1}}$Department of Engineering Cybernetics, Norwegian University of Science and Technology, 7491 Trondheim, Norway.\\
$^{\mathsf{2}}$Delft University of Technology, Delft Center for Systems and Control, 2628 CD
Delft, The Netherlands.}

\ead{\{leif.e.andersson, lars.imsland\}@ntnu.no, \{b.m.doekemeijer, d.c.vanderhoek,j.w.vanwingerden\}@tudelft.nl}

\begin{abstract}
This article investigates the optimization of yaw control inputs of a nine-turbine wind farm. The wind farm is simulated using the high-fidelity simulator SOWFA. The optimization is performed with a modifier adaptation scheme based on Gaussian processes. Modifier adaptation corrects for the mismatch between plant and model and helps to converge to the actual plan optimum. In the case study the modifier adaptation approach is compared with the Bayesian optimization approach. Moreover, the use of two different covariance functions in the Gaussian process regression is discussed. Practical recommendations concerning the data preparation and application of the approach are given. It is shown that both the modifier adaptation and the Bayesian optimization approach can improve the power production with overall smaller yaw misalignments in comparison to the Gaussian wake model.   
\end{abstract}
\section{Introduction}
Wind energy is a core part of the effort to reach a carbon neutral society in Europe and worldwide. Nowadays about $5\%$ of the global energy production is supplied by wind \cite{Veers2019}. However, in the coming years the deployed capacity will increase rapidly in order to meet the high renewable energy targets. The European Commission plans to install \SI{450}{GW} of offshore wind by 2050 \cite{Freeman2019}. Most of the wind turbines will be grouped together and installed near each other to reduce maintenance and deployment costs \cite{Fleming2016}. However, when a wind turbine extracts energy from the wind a \textit{wake} behind the turbine develops. The wake is characterized by a reduced flow velocity and higher turbulence intensity. A turbine operating in a wake of another turbine extracts less energy and experiences higher load variations \cite{Steinbuch1988}. Clearly coordinated wind farm control strategies that dampen the interaction between turbines have the potential to reduce the levelized cost of energy \cite{Boersma2019a} and make wind energy even more competitive. \\
%Currently the most promising wind farm control strategy is wake steering \cite{Wagenaar2012,Park2013,Gebraad2016}. 
%
%The most common solutions to the wind farm control problem are power de-reating control, e.g. \cite{Rotea2014,Munters2016,Duc2019} and wake steering control, e.g. \cite{Wagenaar2012,Park2013,Gebraad2016}. In the former solution the upstream turbines are operated sub-optimally by extracting less energy from the wind than possible. As a result, a smaller velocity reduction is caused in the wind, which can be exploited by the downwind turbines. The target net effect is an increase in the overall energy production. Static power de-rating received mix results in wind tunnel experiments \cite{Campagnolo2016a}, field studies \cite{Schepers2007,Hoek2019} and high fidelity simulations \cite{Annoni2016}. However, recently dynamic power de-rating is studied, which may re-vitalise this wind farm control solution \cite{Munters2018a,Frederik2019}. \\ 
Currently a promising approach to improve the power production with wind farm control is wake steering. A yaw offset is applied to the upstream turbine misaligning the rotor plane with the inflow and deflecting the wake away from the downwind turbine. Significant power gains can be achieved downstream resulting in overall power gains in the wind farm. Moreover, field experiments to prove the effect of wake steering have shown promising results \cite{Fleming2017a,Fleming2019,Howland2019}. \\ 
Model-based wind farm control usually relies on simplified surrogate models to estimate the wake interactions. These models often calculate the steady-state situation and can represent the general behavior of wakes \cite{Barthelmie2013,Annoni2014,Bastankhah2016}. Nonetheless, the complex dynamics of the air flow and the large range of spatial and temporal scales make the development of an accurate surrogate model challenging \cite{Doekemeijer2019a}. Improving and developing new surrogate models is a very active research field \cite{Martinez-Tossas2018,Sun2018,Kabir2020,Ge2019}. Model-free optimization methods were investigated to drive the wind farm iteratively to a plant optimum without the requirement of a possible inaccurate model \cite{Marden2013,Ciri2017}. However, these methods suffer from long convergence times. \\
Modifier-adaptation (MA) is a real-time optimization (RTO) strategy that uses measurement data to correct the cost and constraint functions of the optimization problem directly, and reaches, under suitable assumptions, true plant optimality upon convergence \cite{Marchetti2009}. Consequently, MA can correct the inaccurate surrogate model using plant measurements and provides a viable closed-loop model-based wind farm control solution. \\
Recently, a method was proposed that combines MA with Gaussian process (GP) regression for wind farm optimization \cite{Andersson2020c}. The GP regression is used to correct for the plant-model mismatch. In \cite{Andersson2020} the approach was improved. Instead of identifying the plant-model mismatch of the power production of the wind farm, the plant-model mismatch of the power production of each single turbine was identified with multiple GP regression models. The multiple-input and multiple-output approach needs less data to achieve the same performance and scales better for large wind farms compared to the multiple-input and single-output approach. \\ 
In this article, MA and GP regression (MA-GP) are used to optimize the power production in a wind farm using data generated with the high-fidelity simulator SOWFA \cite{Churchfield2012}. It is shown that the approach can handle noisy and dynamic data. Moreover, the approach achieves satisfactory result with a small data set. \\ 
The article is organized as follows: In section \ref{ref:Method} the methodologies used in the article are introduced. It follows a detailed description of the Case study in section \ref{sec:CaseStudy}. The article ends with a conclusion.
\section{Methodology}\label{ref:Method}
This section gives a brief description of the MA-GP approach. A more detailed description of the MA-GP approach for wind farms can be found in \cite{Andersson2020}. First, the MA approach and objective are introduced in Section \ref{sec:MA}, then the GP regression is described in \ref{sec:GP-R} and finally the MA-GP approach is presented in Section \ref{sec:MAGP}.
\subsection{Modifier Adaptation}\label{sec:MA}
The objective of the modifier adaptation approach is to find the optimal steady-state operation point of the plant. The challenge in real-time optimization is that usually an exact input-output map of the plant is unknown and instead a model approximation is exploited in the optimization. Inevitably the plant-model mismatch introduced by the model can lead to sub-optimal performance of the real-time optimization. MA improves the optimization by correcting the objective and constraint functions of a real-time optimization problem using plant measurements. \\
In context of the wind farm optimization, the model consists of a wake and turbine model. The objective is to maximize the power production and the optimization variables are the turbine yaw angles. \\
A computational bottleneck of the MA approach is the calculation of the plant gradients, which are required in the correction of the objective and constrained functions. The use of GPs was proposed by \cite{AvilaFerreira2018} to overcome this bottleneck.
\subsection{Gaussian process regression}\label{sec:GP-R}
GPs are based on kernel methods \cite{Rasmussen2006}. They aim to describe an unknown function $f:\mathbb{R}^{n_u} \rightarrow \mathbb{R}$ from data. It is assumed that the noisy observation of $f(\cdot)$ are given by 
\begin{equation}
    y = f(\mathbf{u}) + \nu \label{eq:measurement}, 
\end{equation}
where the noise $\nu$ is Gaussian with zeros mean and variance $\sigma^2_{\nu}$ and $\bu$ is the input, which is assumed to follow a multivariate Gaussian distribution. A covariance function and mean function determine the smoothness and continuity properties of the underlying function \cite{Snelson2006}. In this article a zero mean function and the automatic relevance (ARD) squared-exponential (SE) covariance function are chosen as a default choice: 
\begin{equation}
    k(\bu_i,\bu_j) = \sigma_f^2 \exp\left(-\frac{1}{2}\btau_{ij}^T \bLambda^{-1} \btau_{ij} \right) \label{eq:SE_eq},
\end{equation}
where $\sigma_f^2$ is the covariance magnitude, $\bLambda = \text{diag}(\lambda_1^{2},\dots,\lambda_{n_u}^{2})$ is a scaling matrix and $\btau_{ij} = \bu_i-\bu_j$. 
Alternatively, the Mat\'{e}rn $\nicefrac{5}{2}$ covariance function (M$\nicefrac{5}{2}$) is chosen,
\begin{equation}
    k(\bu_i,\bu_j) = \sigma_f^2 \left(1+\sqrt{5 \btau_{ij}^T \bLambda^{-1} \btau_{ij}} + \frac{5}{3} \btau_{ij}^T \bLambda^{-1} \btau_{ij}  \right) \exp\left( -\sqrt{5 \btau_{ij}^T \bLambda^{-1} \btau_{ij}} \right) \label{eq:Matern_eq}, 
\end{equation}
which is less smooth and better suited for practical optimization problems \cite{Snoek2012}. 
\\The hyperparameters $\bpsi := [\sigma_f,\sigma_{\nu},\lambda_1,\ldots,\lambda_{n_{\mathbf{u}}}]^T$ are inferred from data maximizing the log marginal likelihood. The output of a prediction with a GP is a mean and variance value. Consequently, it is possible to evaluate the uncertainty of an operating point and even include this information in the optimization, which is a desired property.  
\subsection{Modifier Adaptation with Gaussian processes}\label{sec:MAGP}
\begin{figure*}[ht]
        \centering			\includegraphics[width=0.70\textwidth]{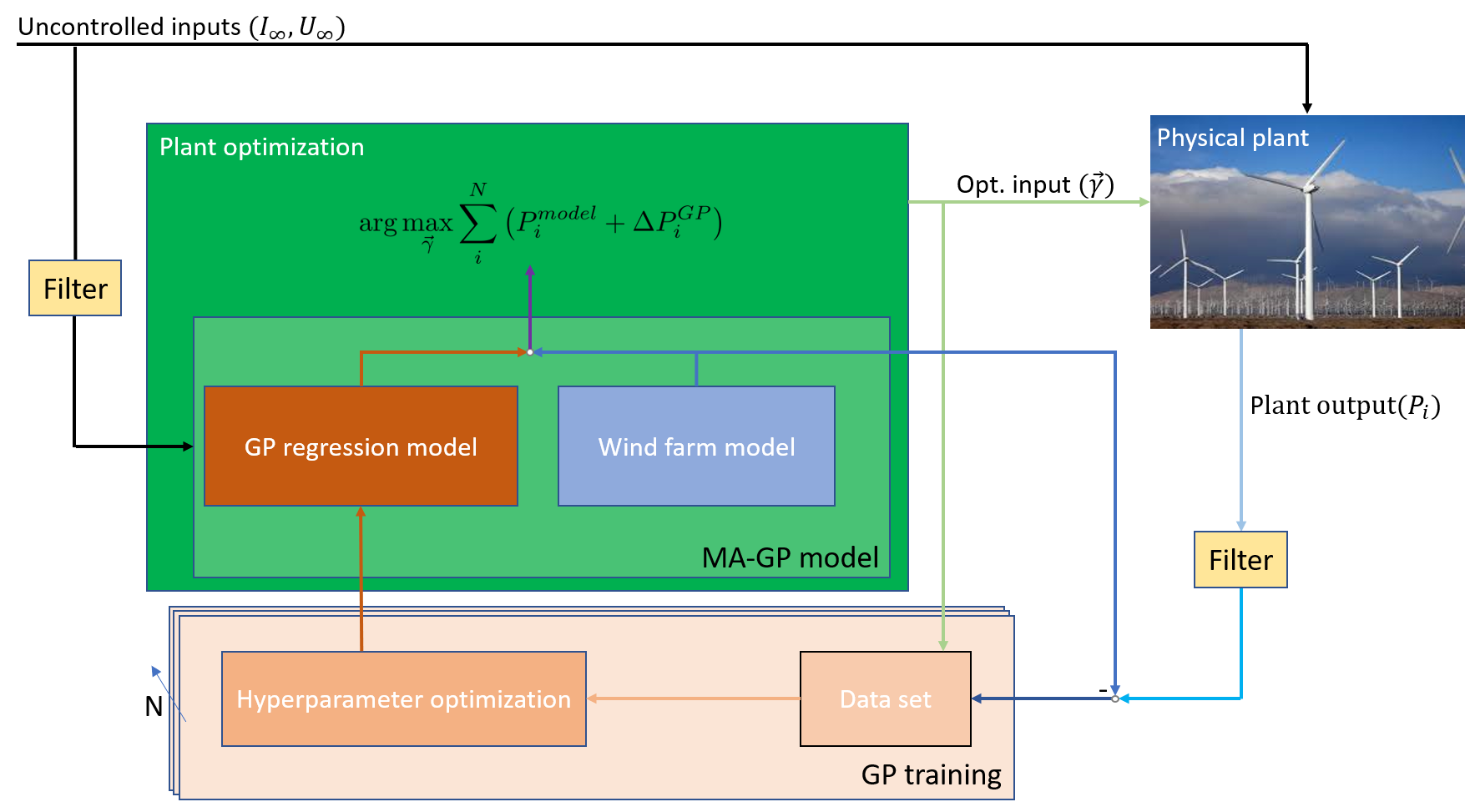}
				\caption{The basic idea of the MA-GP scheme for a wind farm  \protect\footnotemark. }
				\label{fig:MA-GP_Fig}
		\end{figure*}

The idea is to identify the difference between plant and model with a GP using plant measurements. \\
In context of this article the GP regression model creates an input-output map of the control inputs to the plant-model mismatch of the power production for each wind turbine. In fact, GP regression models are used to correct the power outputs of the approximate model. The resulting MA-GP model is used in the optimization to compute optimal yaw inputs for each wind turbine. The inputs and the difference between the measured and estimated outputs of plant and model, respectively, are used to update the data set and the hyperparameters of each GP regression model. The power measurements of the plant are filtered with a moving average filter (Fig. \ref{fig:MA-GP_Fig}). Filtering is necessary since the steady-state behavior of the plant is represented by the approximate model.
		\footnotetext{The wind farm picture is by Erik Wilde from Berkeley, CA, USA \url{https://www.flickr.com/photos/dret/24110028330/}, \textit{Wind turbines in southern California 2016}, \url{https://creativecommons.org/licenses/by-sa/2.0/legalcode }}
\subsection{Bayesian Optimization}
The MA-GP approach is compared with a GP regression approach that identifies the plant without additional model (Bayesian optimisation (BO) approach). The only difference between the MA-GP and the BO approaches is the inclusion of the \textit{Wind farm model} in the MA-GP approach. Consequently, the BO approach identifies the power production of each turbine directly instead of the plant-model mismatch. 
\section{Case study}\label{sec:CaseStudy}
\begin{figure*}[ht]
        \centering
				\includegraphics[width=0.40\textwidth]{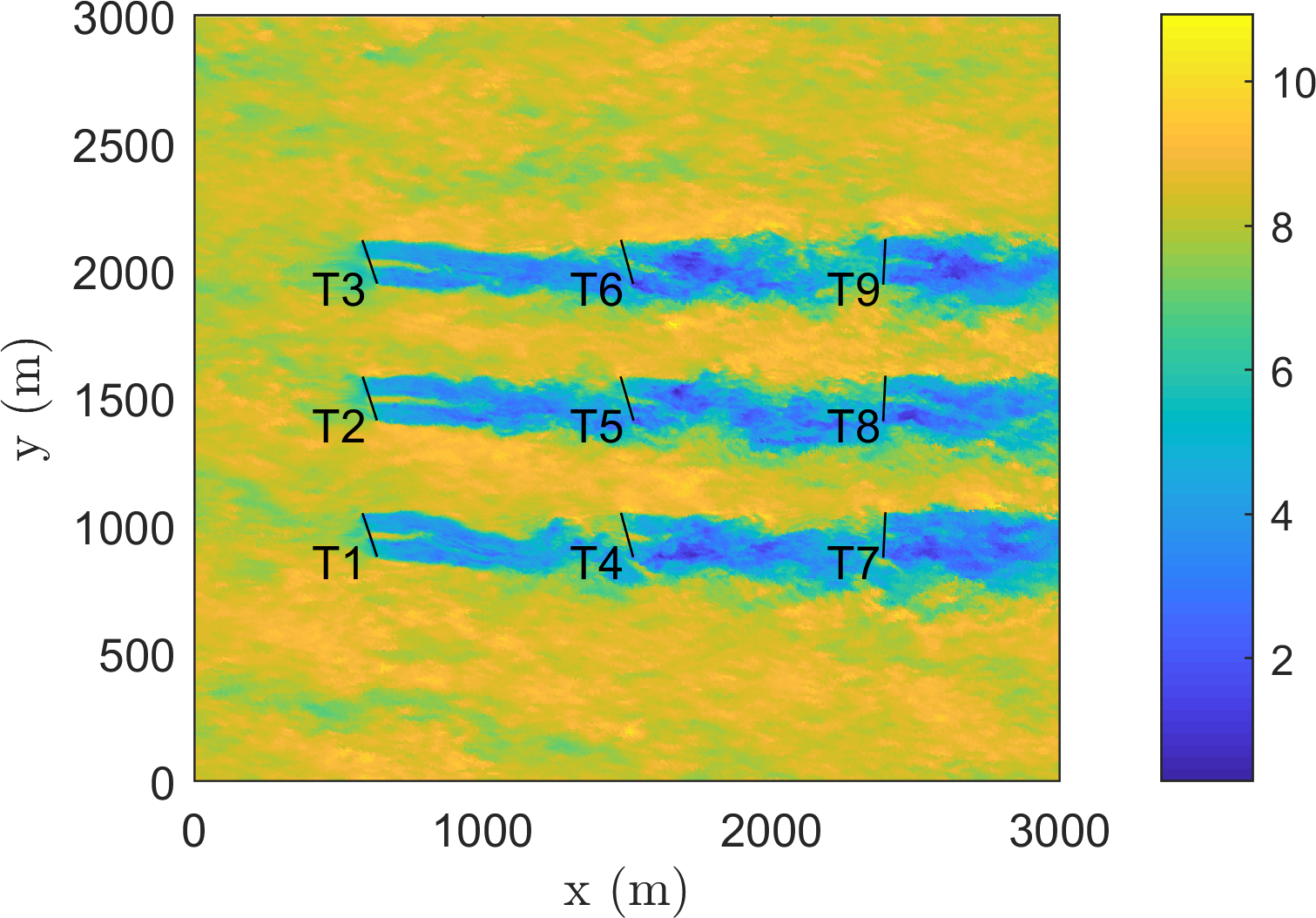}
				\caption{Top view of a snap shot on the wind farm simulated with SOWFA \cite{Doekemeijer2019a}.}
				\label{fig:SOWFAPlant}
		\end{figure*}

The proposed MA-GP approach is tested in a high-fidelity simulation for a wind farm with nine DTU \SI{10}{MW} turbines (Fig. \ref{fig:SOWFAPlant}). It is the same setup as presented in \cite{Doekemeijer2019a}. The simulation settings are displayed in Tab. \ref{tab:SOWFASetup}.  \\ 
\begin{table}[h!]
\centering
		\renewcommand{\arraystretch}{1.1}
		\captionsetup{width=.98\textwidth}
		\caption{Simulation settings for SOWFA.}
		\label{tab:SOWFASetup}
    \scriptsize
\begin{tabulary}{0.9\columnwidth} {l l}
Variable & Value \\\hline 
Timestep & \SI{0.2}{s} \\ 
Cell size (near rotor) & \SI{2.5}{m} $\times$ \SI{2.5}{m} $\times$ \SI{2.5}{m} \\ 
Cell size (outer region) & \SI{10.0}{m} $\times$ \SI{10.0}{m} $\times$ \SI{10.0}{m} \\ 
Blade epsilon, $\varepsilon$ & \SI{5.0}{m} \\ 
Turbine layout & \num{3} $\times$ \num{3} turbines at \SI{5}{D} $\times$ \SI{3}{D} spacing \\ Freestream wind speed, $U_\infty$ & \SI{8.0}{m/s} \\ 
Freestream turbulence intensity, $I_\infty$ & \SI{5.0}{\%} \\ 
Freestream wind direction, $\phi$ & \SI{0.0}{rad} 
\end{tabulary}
\end{table}
\subsection{Data preparation}
The generator power at each turbine is measured. A moving average filter with a time horizon of five minutes is used to filter the measurements. Each operating point (OP) is simulated for \SI{10}{min}. The transient of the step response is excluded by removing the first five minutes after a change of an OP from the data. Consequently, for each OP one power measurement was included into the training sets of the GPs. The initial training set consisted of 20 OPs. However, several OPs were similar or even the same such that effectively only 15 different OPs were available. Nonetheless, equal and similar operating points can be used to estimate the noise of the process, so they contain some information. The lower and upper boundaries of the yaw misalignment of the three turbine rows are $\gamma_l = [-\ang{25},-\ang{20},-\ang{7}]^T$ and $\gamma_u = [\ang{18},\ang{10},\ang{5}]^T$, respectively. A larger variation for the upstream turbines was allowed since it is expected that the optimal OP requires larger yaw misalignment of the upwind turbines.
%\footnotetext{The larger variance in the counter-clockwise rotation was allowed by mistake due to different yaw angle definition in the Gaussian wake model and SOWFA.}.  
%
\subsection{Surrogate model}
The wake is approximated by the Gaussian wake model \cite{Bastankhah2016} and the wind turbines by the actuator disc theory \cite{Burton2011}. The model adaptation method proposed by \cite{Doekemeijer2019a} is applied to improve the power estimates of the surrogate model. The ambient wind velocity and turbulence intensity are estimated by minimizing the difference between the time averaged power measurements of SOWFA and the surrogate model.
\subsection{Results - Upstream turbines}\label{sec:Results_Row}
\begin{figure*}[t]
        \centering
        \begin{subfigure}[t]{0.49\columnwidth}
        \centering
				\includegraphics[width=0.925\columnwidth]{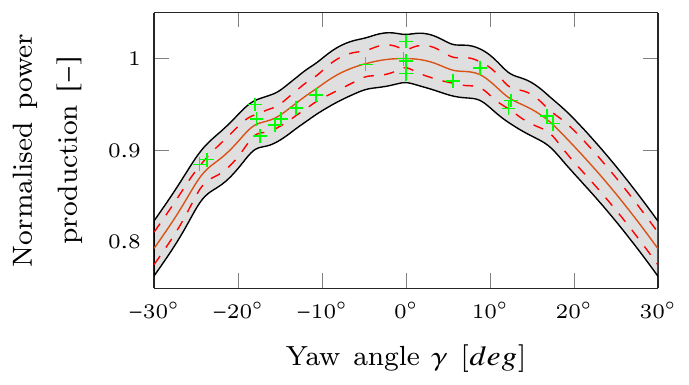}
    		\captionsetup{width=.975\linewidth}
    		\caption{Turb. 1; 20 training points; MA-GP.}
    		\label{fig:Turbine1_20training_MAGP}
    	\end{subfigure}	
    	\begin{subfigure}[t]{0.49\columnwidth}
        \centering
				\includegraphics[width=0.8\columnwidth]{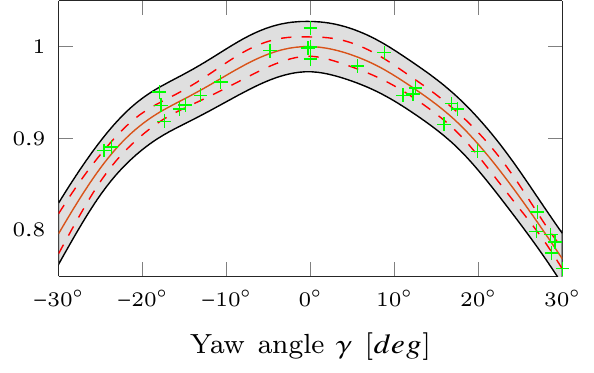}
    		\captionsetup{width=.975\linewidth}
    		\caption{Turb. 1; 29 training points; MA-GP.}
    		\label{fig:Turbine1_29training_MAGP}
    	\end{subfigure}	
    	
    	\begin{subfigure}[t]{0.49\columnwidth}
        \centering
					\includegraphics[width=0.925\columnwidth]{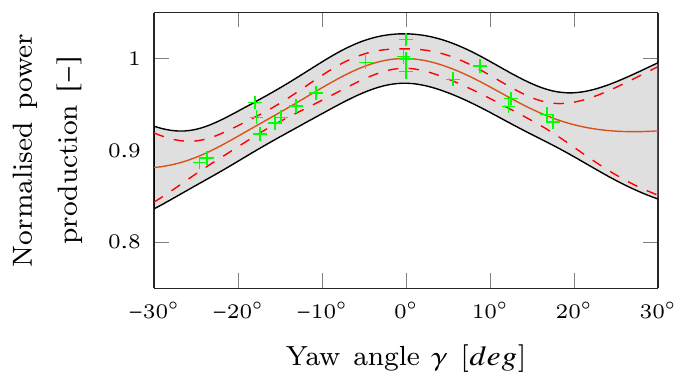}
    		\captionsetup{width=.975\linewidth}
    		\caption{Turb. 1; 20 training points; BO.}
    		\label{fig:Turbine1_20training_GP}
    	\end{subfigure}	
    	\begin{subfigure}[t]{0.49\columnwidth}
        \centering
					\includegraphics[width=0.8\columnwidth]{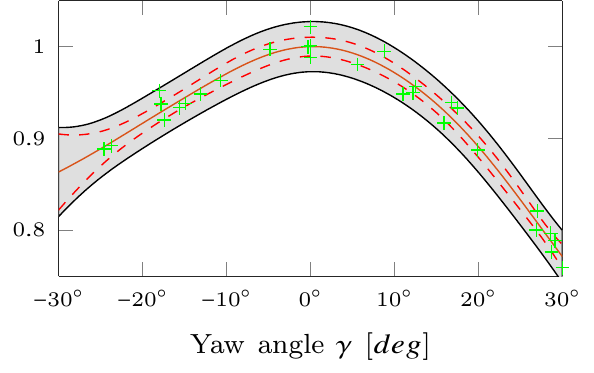}
    		\captionsetup{width=.975\linewidth}
    		\caption{Turb. 1; 29 training points; BO.}
    		\label{fig:Turbine1_29training_GP}
    	\end{subfigure}	
    	
    	\begin{subfigure}[t]{0.49\columnwidth}
        \centering
				\includegraphics[width=0.925\columnwidth]{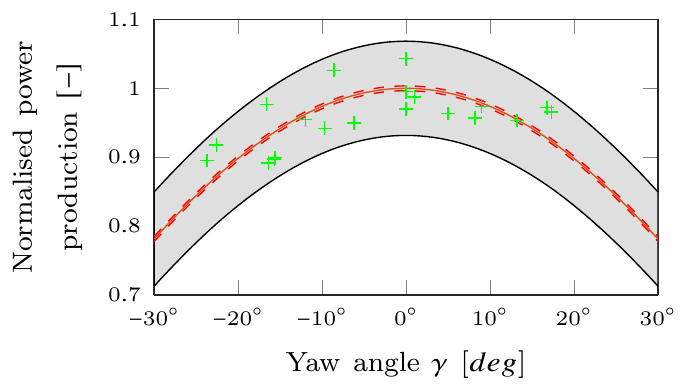}
    		\captionsetup{width=.975\linewidth}
    		\caption{Turb. 3; 20 training point; MA-GP.}
    		\label{fig:Turbine3_20training_MAGP}
    	\end{subfigure}	
    	\begin{subfigure}[t]{0.49\columnwidth}
        \centering
				\includegraphics[width=0.8\columnwidth]{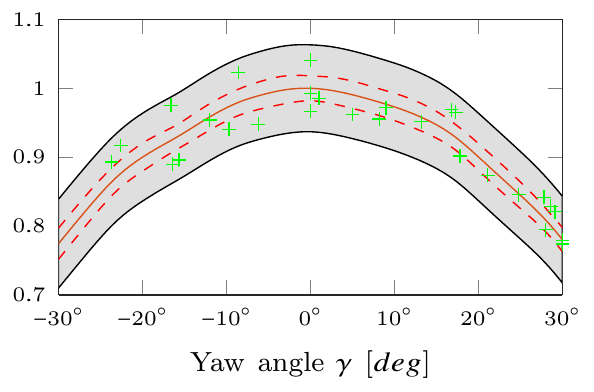}
    		\captionsetup{width=.975\linewidth}
    		\caption{Turb. 3; 29 training point; MA-GP.}
    		\label{fig:Turbine3_29training_MAGP}
    	\end{subfigure}	
    	
    	\begin{subfigure}[t]{0.49\columnwidth}
        \centering
					\includegraphics[width=0.925\columnwidth]{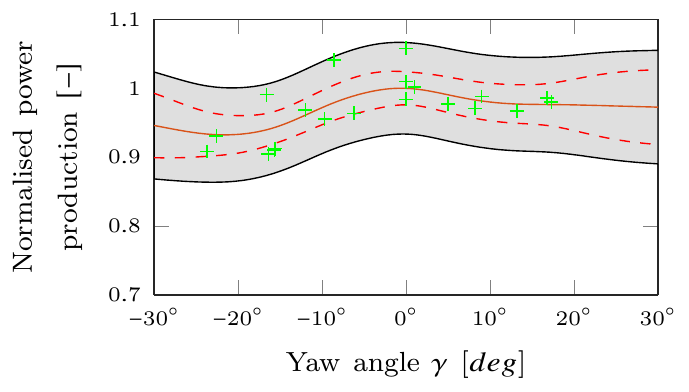}
    		\captionsetup{width=.975\linewidth}
    		\caption{Turb. 3; 20 training point; BO.}
    		\label{fig:Turbine3_20training_GP}
    	\end{subfigure}	
    	\begin{subfigure}[t]{0.49\columnwidth}
        \centering
					\includegraphics[width=0.8\columnwidth]{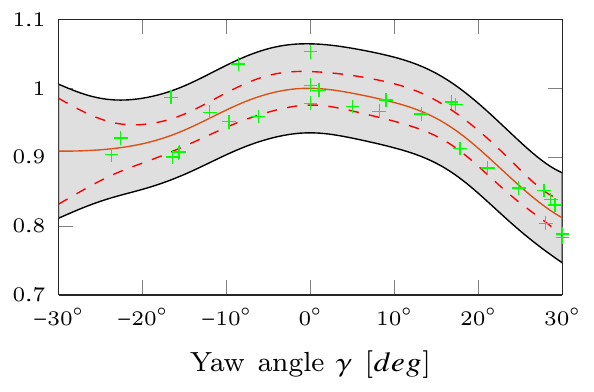}
    		\captionsetup{width=.975\linewidth}
    		\caption{Turb. 3; 29 training point; BO.}
    		\label{fig:Turbine3_29training_GP}
    	\end{subfigure}	
    	\caption{The normalized power production of the turbine 1 and turbine 3 identified with the MA-GP and BO approaches. The orange line represents the mean value, the red dashed line indicates the \SI{95}{\%} certainty range of the model and the grey region indicates the \SI{95}{\%} certainty range of model plus measurement noise. The green crosses represent (filtered) measurement points.}
    		\label{fig:Results_1}
\end{figure*}

It is difficult to visualize the objective function of a plant with nine control inputs. Therefore, the model identification of the upstream turbines, using MA-GP and BO, is investigated and compared (Fig. \ref{fig:Results_1}).  
The power production of turbine 1 with the MA-GP approach is well captured (Fig. \ref{fig:Turbine1_20training_MAGP}). It aligns with the expected values from SOWFA \cite{Doekemeijer2019a}. The BO approach (Fig. \ref{fig:Turbine1_20training_GP}) gives similar predictions in the vicinity of the measurement points. At the boundaries (large absolute yaw angles) the mean values of the power production flatten. For even larger yaw angles the power production even increases again. The method converges to about the mean value of the data points and the variance increases considerably. For the MA-GP approach this behavior cannot be observed. For the large yaw angles a constant bias is corrected and the variance converges quickly to a constant value. \\ 
For turbine 3 the uncertainty in the prediction of the power production is larger. For the MA-GP approach (Fig. \ref{fig:Turbine3_20training_MAGP}) an almost constant bias correction independent of the yaw angle is applied to the turbine. For the BO approach (Fig. \ref{fig:Turbine3_20training_GP}) the power production is almost independent of the yaw angle and rather flat. It follows the mean of the data set. The reason is the large noise of the measured power production. The power production of the yawed case is similar and sometimes even larger than for the non-yawed case. Consequently, the GP regression identifies a large measurement noise. The variation in the data is then mainly explained by the measurement noise without strong variation in the estimate of the mean value. \\ 
The figures in the right column (Fig. \ref{fig:Turbine1_29training_MAGP} - \ref{fig:Turbine3_29training_GP}) show the results with some more data points. The turbines are operated with a yaw angle between \angRange{20}{30}. The MA-GP approach extrapolated quite well. Only the power production at turbine 1 is slightly corrected downward. The variance estimate at turbine 3 is corrected. The model uncertainty is increased while the measurement noise estimate is reduced. The previously extrapolated results of the BO approach for large yaw angles is corrected significantly.
\subsection{Practical recommendations}
\begin{figure}[t]
        \centering
				\includegraphics[width=0.95\columnwidth]{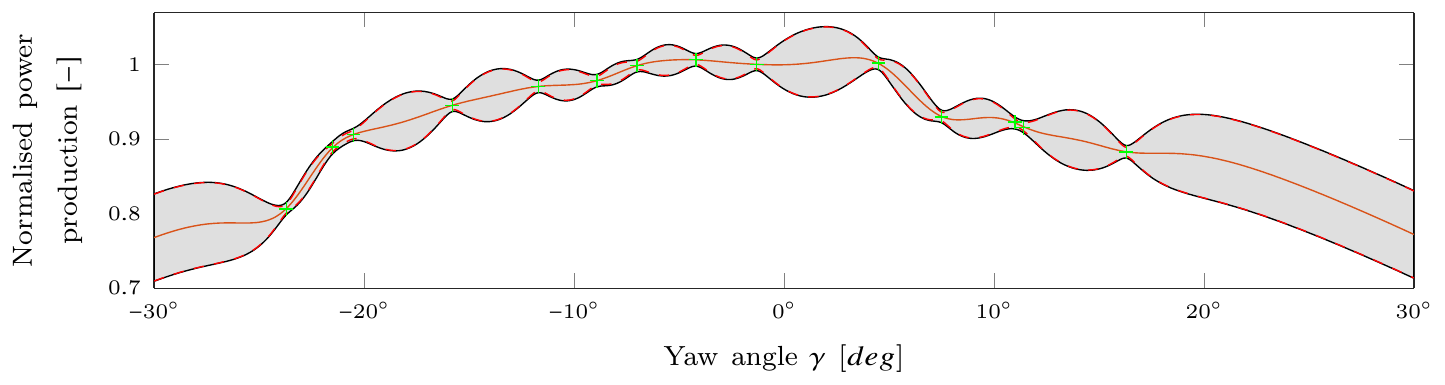}
    		\captionsetup{width=.95\linewidth}
    		\caption{GP regression with overfitting.}
    		\label{fig:Results_2}
\end{figure}

Overfitting can be an issue if only a few data points are used in the identification, cf. Fig. \ref{fig:Results_2}. Here, the measurement noise $\sigma_v$ and the length scales $\lambda_1,\dots,\lambda_{n_u}$ are estimated to be very small. Therefore, the data points are fitted almost perfectly by the GP giving the mean estimate a not smooth behavior. Including some similar operating points (with different outputs), which help to estimate the measurement noise, can prevent this phenomenon. Another solution is to set a lower bound on the hyperparameters in the hyperparameter optimization.   \\
As expected, the M$\nicefrac{5}{2}$ covariance function is less smooth and certain about the estimates than the SE covariance function. With few data points the M$\nicefrac{5}{2}$ covariance function is more prone to overfitting. Otherwise, both covariance functions show similar results in estimating the power production of the first turbine row. 
\subsection{Results - Optimization of plant power production}
\begin{figure}[t]
        \centering
				\includegraphics[width=0.95\columnwidth]{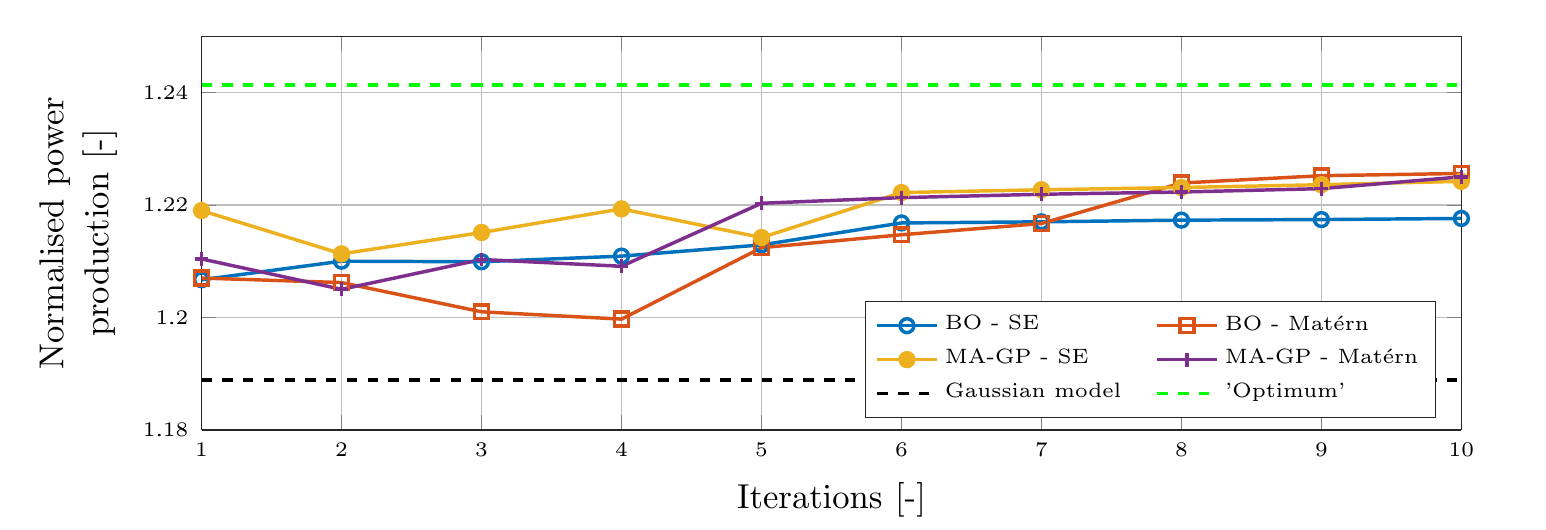}
    		\captionsetup{width=.95\linewidth}
    		\caption{Operating point normalized by the greedy power production over the ten iterations. The lower dashed line shows the optimal power production of the Gaussian wake model. The upper dashed line shows the optimal power production of the scaled three-turbine plant. It is the best power production reached in the SOWFA simulations.}
    		\label{fig:OptIteration1_3}
\end{figure}

The results of the optimization over several iterations will be discussed. The algorithms are initialized with 20 data points. The optimal operating points are calculated using the MA-GP and BO approach. Subsequently, the optimal operating point is applied in the SOWFA simulation to get a new power measurement. \\
Comparing the two covariances and the MA-GP and BO algorithms would have led to many time-consuming SOWFA simulations. Therefore, only for two methods the optimized yaw angles were sent to SOWFA. For the other approaches a \textit{quasi-SOWFA} plant model was created. A three-turbine wind farm was simulated with a GP model including all available data. This was achieved by scaling the power production of each SOWFA run with the power production of the greedy setup and the same wind input. It removed the variance in the power production. Afterwards the turbine rows can be separated, and the power production of a three- turbine wind farm can be optimized or predicted using a larger dataset. With this trick we were able to represent \textit{quasi-SOWFA} predictions. The power predictions in the resulting three-turbine plant is very accurate for each turbine row of the nine-turbine wind farm. \\
The results of ten iterations using the different algorithms and the quasi-SOWFA predictions as a plant model are shown in Fig. \ref{fig:OptIteration1_3}. After ten iterations the power production increase ranges between \SIrange{21.8}{22.5}{\%}. \\ 
The optimal yaw misalignment is about $\vec{\gamma} = [\ang{30},\ang{12},\ang{0}]^T$ for each of the rows and a power gain of about \SI{24}{\%} (in actual SOWFA runs) can be achieved. This optimum was found with the scaled three turbine plant. The yaw misalignment of the second turbine varies between \angRange{11}{26} for both the BO and MA-GP approach, which indicates that the total power production of the plant might be less sensitive to the yaw misalignment of the first downwind turbine. The power production at the optimum is more distributed over all turbines. The second and third turbine produce almost the same power. \\
At the initialization the MA-GP approach is superior to the BO approach. After ten iterations the differences between the three best methods is within the expected variance of the power production estimate. In addition, the absolute yaw misalignment is similar for both approaches. \\ 
Including an exploration term in the objective function may help to converge closer to the optimum. In comparison to the optimization using solely the wind plant model the BO and MA-GP approaches increase the power production about \SIrange{2}{3}{\%} with overall smaller yaw misalignments. Even at the initialization the power production is about \SIrange{1.5}{2.5}{\%} higher. \\
Even though both approaches have a similar performance it is recommended to use the MA-GP approach. Without constraints, the BO approach tends to very large yaw angles in the first turbine row because of the behavior described in Sec. \ref{sec:Results_Row}. Moreover, a similar behavior can be observed for the last turbine row, which tend also sometimes to large yaw angles. These issues cannot be observed in the MA-GP approach. 
Solutions to prevent this problem are: 
\begin{itemize}
    \item Filtering of the optimization results: 
    \begin{equation}
        \vec{\gamma}_{k+1} = \vec{\gamma}_{k} + \kappa (\vec{\gamma}_{k+1}^* - \vec{\gamma}_{k}),\end{equation}
        where $\vec{\gamma}^*$ are the optimal yaw angles and $\kappa$ a filter variable.  
    \item Constrain the optimization variables so extrapolation is avoided, 
    \item Include the variance in the objective function 
    \begin{equation}
        \vec{\gamma}_{k+1}^* = \text{arg}\max_{\vec{\gamma}} \left( \boldsymbol{\mu}^2 - \tau \boldsymbol{\Sigma} \right),  \end{equation} where $\boldsymbol{\mu}$ represents the power production of the wind farm, $\boldsymbol{\Sigma}$ the variance in the power production estimate and $\tau$ a weighting factor.  
    \item Use the trust-region approach proposed by \cite{RioChanona2019}
\end{itemize}
If the applied constraints allow only a small feasible region they should be relaxed if the algorithm converged to a boundary. Furthermore, the penalization of the variance should also be relaxed once the data set grows to allow exploration. Otherwise, it may happen that the algorithm converges to the most certain point. On the other hand, the variance only increases strongly for the BO approach if the approach extrapolates. Therefore, it can be expected that the variance term has not a strong influence on the objective function if weighted well.\\  
%It was observed in simulations, where the GP with the full data set was used to simulate the plant, that the BO approach tends to slightly higher power production. It seems being unbiased by a model can have a positive effect. The behavior has to be investigated in further simulations. 
% 
\subsubsection{Practical recommendations regarding the filter horizon}
\begin{figure*}[t]
        \centering
        \begin{subfigure}[t]{0.49\columnwidth}
        \centering
				\includegraphics[width=0.925\columnwidth]{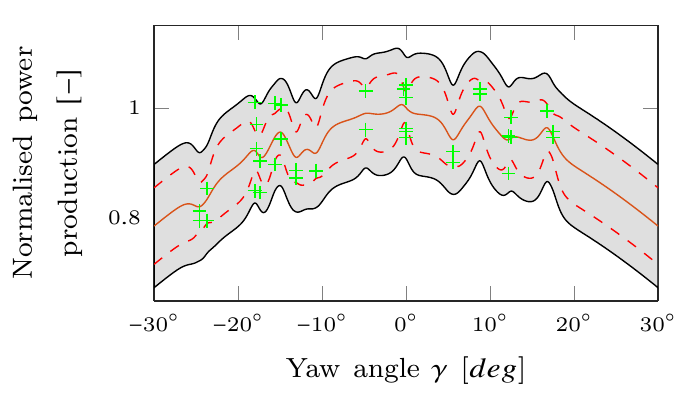}
    		\captionsetup{width=.95\linewidth}
    		\caption{40 training points; MA-GP.}
    		\label{fig:Turbine1_20train_MAGP_2.5minfilter}
    	\end{subfigure}	
    	\begin{subfigure}[t]{0.49\columnwidth}
        \centering
				\includegraphics[width=0.82\columnwidth]{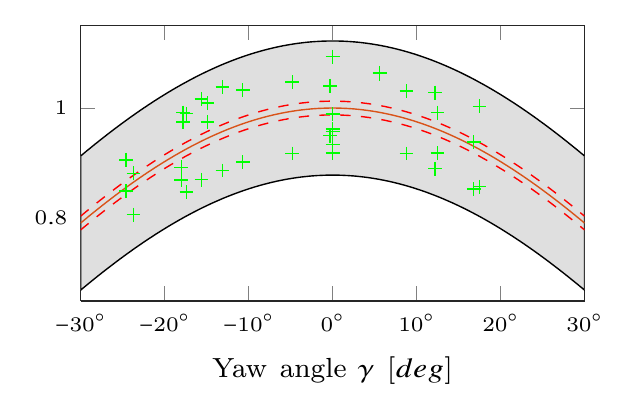}
    		\captionsetup{width=.95\linewidth}
    		\caption{40 training points; MA-GP; $\sigma_v$ constrained.}
    		\label{fig:Turbine1_20train_MAGP_2.5minfilter_Const-1}
    	\end{subfigure}	
    	
    	\begin{subfigure}[t]{0.49\columnwidth}
        \centering
					\includegraphics[width=0.925\columnwidth]{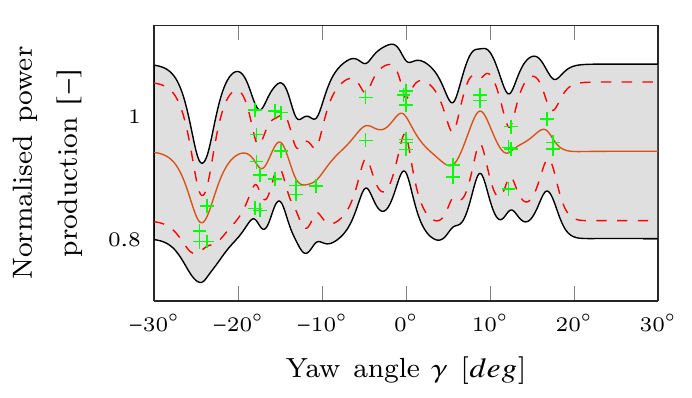}
    		\captionsetup{width=.95\linewidth}
    		\caption{40 training points; BO.}
    		\label{fig:Turbine1_20train_GP_2.5minfilterP}
    	\end{subfigure}	
    	\begin{subfigure}[t]{0.49\columnwidth}
        \centering
					\includegraphics[width=0.82\columnwidth]{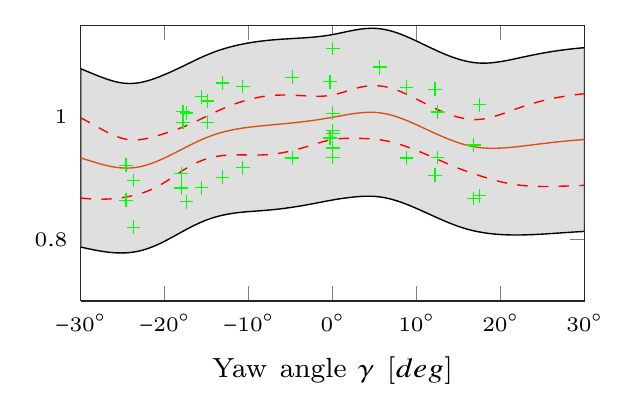}
    		\captionsetup{width=.95\linewidth}
    		\caption{40 training points; BO; $\sigma_v$ constrained.}
    		\label{fig:Turbine1_20train_GP_2.5minfilter_Const-1}
    	\end{subfigure}	
    	\caption{The normalized power production of turbine 1 and turbine 3 identified with the MA-GP and BO approaches using a filter horizon of \SI{2.5}{min}. The left column shows the results with unconstrained hyperparameters. The right column shows the result with constrained hyperparameters.}
    		\label{fig:Results_3}
\end{figure*}

Previously a five-minute horizon was used to filter the power outputs of the turbines. A shorter filter horizon increases the variance of the data. We like to present some preliminary observation if a filter horizon of \SI{2.5}{min} is used, but the data in the GP doubled.  We observe a tendency to overfitting in the GP regressions. Moreover, in the BO approach the power production becomes less sensitive to yaw angle changes (Fig. \ref{fig:Turbine1_20train_MAGP_2.5minfilter_Const-1}). \\
Again, the M$\nicefrac{5}{2}$ covariance function is more prone to overfitting than the SE covariance. Overfitting can be reduced constraining the hyperparameters or simply by a larger dataset with different operating points.  
The optimization of the power production using the initial data set of \SI{40} data points is negatively affected by the shorter filter horizon for both the BO and MA-GP approaches. Both approaches show similar performance. Interestingly, the SE covariance behaves superior to the M$\nicefrac{5}{2}$ covariance function. It is probably a direct consequence of the smoother property of the SE covariance. If the noise term $\sigma_v$ is lower bounded the performance of the MA-GP approach increases. For the BO approach the results are inconclusive. Both tendencies were observed. It can be observed that smaller yaw misalignments are used, which indicates that constraining the $\sigma_v$ might lead to a more robust approach in case of very noisy data. 
A more in-depth study using Monte-Carlo simulations must be performed to draw more conclusive results.
\section{Discussion $\&$ Conclusion}
In this article the MA-GP approach proposed by \cite{Andersson2020c} is applied to a nine-turbine wind farm simulated with SOWFA. Alternatively, to the MA-GP approach the BO approach is tested, which uses no a priori plant model. For both approaches the multiple-input multiple output identification scheme proposed by \cite{Andersson2020} is applied. Even though a relatively small data set is used as initial training set, which contained a large amount of sub-optimal counterclockwise yaw settings, the algorithms can converge quickly to a near optimal solution. For the reported power gains, it must be kept in mind that the inflow in SOWFA is quasi-static, the free-stream turbulence intensity is relatively small and the plant alignment with the inflow is favorable for wake steering. \\
The MA-GP and BO approaches can improve the power production compared to the Gaussian wake model. Moreover, the expected better performance for clockwise yaw misalignment is identified and the surrogate model improved. A clear recommendation for either the MA-GP or the BO approach cannot be given. Both approaches showed similar performance. The MA-GP approach might be better suited for small data sets and large wind farms because of its better extrapolation property. Even though both covariance functions show similar performance the smoothness of the SE covariance seems to be a favorable feature making it more user-friendly. \\
In future work the recommendations and observations pointed out in the article should be verified in Monto-Carlo simulations. In such a setup the most accurate GP model will be used as the plant model to avoid time-consuming SOWFA simulations.
%
%
%%%%%%%%%%%%%%%%%%%%%%%%%%%%%%%%%%%%%%%%%%%%%%%%%%%%%%%%%%%%%%%%%%%%%%%%%%%%%%%%
\section*{References}
\balance
\small
\providecommand{\newblock}{}

% that's all folks
\end{document}